\documentclass[aps,prl,twocolumn,groupedaddress,showpacs]{revtex4}

\usepackage{graphicx}

\begin{document}
\title{Coreless vortex formation in a spinor Bose-Einstein condensate}

\author{A.E. Leanhardt}
\author{Y. Shin}
\author{D. Kielpinski}
\author{D.E. Pritchard}
\author{W. Ketterle}

\homepage{http://cua.mit.edu/ketterle_group/}

\affiliation{Department of Physics, MIT-Harvard Center for
Ultracold Atoms, and Research Laboratory of Electronics,
Massachusetts Institute of Technology, Cambridge, Massachusetts,
02139}

\date{\today}

\begin{abstract}

Coreless vortices were phase-imprinted in a spinor Bose-Einstein
condensate. The three-component order parameter of $F=1$ sodium
condensates held in a Ioffe-Pritchard magnetic trap was
manipulated by adiabatically reducing the magnetic bias field
along the trap axis to zero.  This distributed the condensate
population across its three spin states and created a spin
texture. Each spin state acquired a different phase winding which
caused the spin components to separate radially.

\end{abstract}


\maketitle

Spin textures play a central role in describing the physics of
elementary particles~\cite{SKY61}, liquid
$^3$He-A~\cite{MEH76,ANT77,BEK00}, the quantum Hall
effect~\cite{LEK90}, and gaseous Bose-Einstein
condensates~\cite{HO98,OHM98,YIP99,KHS01nature,KHS01pra}.
Topological defects vary between superfluid systems described by
scalar and vector order parameters.  In spin-less or
spin-polarized condensates, line defects such as vortices have
cores where the density of condensed particles is necessarily zero
to keep the order parameter
single-valued~\cite{YGP79,MCW00,ARV01}. However, in condensates
with an internal, spin degree-of-freedom, coreless vortices exist
as spin textures~\cite{BEK00,MAH99}.  Such structures are referred
to as skyrmions (Anderson-Toulouse vortices~\cite{ANT77}) or
merons (half-skyrmions, Mermin-Ho vortices~\cite{MEH76}) depending
on the boundary conditions of the system.

In this Letter, we study spin textures in a Bose-Einstein
condensate (BEC).  Coreless vortices were created in $F=1$ spinor
condensates held in a Ioffe-Pritchard magnetic trap by
adiabatically reducing the magnetic bias field along the trap axis
to zero.  This continuously transformed the initially
spin-polarized condensate into a coherent superposition of three
spin states, each with a different phase winding.  The resulting
angular momentum per particle varied between spin states and the
condensate evolved such that states with more angular momentum per
particle circulated around states with less angular momentum per
particle. Thus, the condensate had a net axial magnetization that
varied with radial position.  Previous work on vortices in a
two-component system used laser, microwave, and radio frequency
fields to spatially and temporally control the interconversion
between components~\cite{MAH99}.  However, without these applied
fields the two-components evolved independently as distinguishable
fluids. In our work, the spin states can freely interconvert at
all points in space and time such that the spin texture would
continually heal itself even in the presence of state-dependent
losses.

In cylindrical coordinates, the spin$-F$ condensate wavefunction
can be written as $|\Psi(r,\phi,z)\rangle = \sqrt{n(r,\phi,z)}
|\zeta(r,\phi,z)\rangle$, where $n$ is the atomic number density
and the $2F+1$ component spinor $|\zeta\rangle = \sum_{m_z=-F}^F
\zeta_{m_z}|F,m_z\rangle,\ |\langle\zeta|\zeta\rangle|^2=1$
describes a spin texture.  A Ioffe-Pritchard magnetic trap
consists of an axial bias field (with curvature) and a
two-dimensional quadrupole field in the orthogonal
plane~\cite{GIT62,DEP83}:
\begin{equation}
\label{e:ioffe}
    \vec{B}(r,\phi,z) = B_z \hat{z} + B'r(\cos (2\phi)\ \hat{r} - \sin (2\phi)\ \hat{\phi}),
\end{equation}
where $B'$ is the radial magnetic field gradient and quadratic
terms have been neglected.  For a condensate of radial extent $R$
confined in a Ioffe-Pritchard magnetic trap with $B_z \gg B'R
> 0$, $|\zeta\rangle = |F,m_z = m_F\rangle$,
where $m_z$ and $m_F$ are the projection of the atomic spin along
the z-axis and local magnetic field direction, respectively.
Adiabatically ramping $B_z$ from $B_z \gg B'R > 0$ to zero rotates
the atomic spin about the position-dependent axis $\hat{n}(\phi) =
\sin \phi\ \hat{x} + \cos \phi\ \hat{y}$, and drives the
transition $|F,m_z=m_F\rangle \rightarrow
\sum_{m_z=-|m_F|}^{|m_F|} \zeta_{m_z} \exp (i(m_z-m_F)\phi)\
|F,m_z\rangle$~\cite{INO00,LGC02}.  Thus, the condensate
population is distributed across $2|m_F|+1$ spin states with each
acquiring a different topological phase factor and angular
momentum per particle due to the variation of Berry's phase with
magnetic quantum number~\cite{BER84}.

\begin{figure}
\begin{center}
\includegraphics{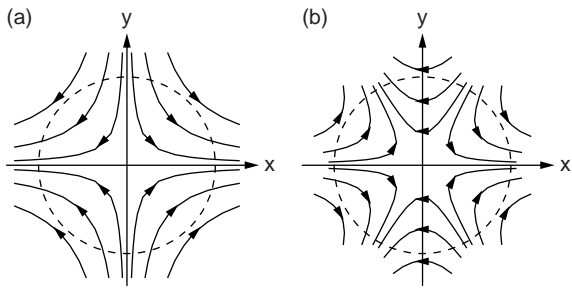}
\caption{Planar spin textures.  Spins aligned with two-dimensional
(a) quadrupole and (b) hexapole magnetic fields produce textures
with winding numbers $-1$ and $-2$, respectively.  In general,
spins aligned with a two-dimensional axisymmetric multipole
magnetic field, $\vec{B} = A r^{q-1}(\cos(q\phi)\ \hat{r} -
\sin(q\phi)\ \hat{\phi}),\ q>0$, produce a texture with winding
number $-(q-1)$.  Counter-clockwise traversal of the dashed
contours in (a) and (b) leads to clockwise (negative) spin
rotation, with the winding number defined as the integer number of
revolutions made by the spin vector while circumnavigating the
singularity at the origin~\cite{MER79}.\label{f:texture}}
\end{center}
\end{figure}

The condensate remains in the state $|F,m_F\rangle$ with respect
to the local magnetic field provided the local Zeeman energy,
$\sim g_F \mu_B ( B_z^2+(B'r)^2)^{1/2}$, dominates the local
kinetic energy associated with the spin texture, $\sim
\hbar^2/mr^2$, where $g_F$ is the Land\'e g-factor, $\mu_B$ is the
Bohr magneton, and $m$ is the atomic mass.  For $B_z = 0$, atomic
spins aligned with the quadrupole magnetic field produce the
planar spin texture in Fig.~\ref{f:texture}(a).  However, the
infinite kinetic energy associated with the wavefunction
singularity at $r=0$ creates a non-planar spin texture over a disc
of radius $\sim (\hbar^2 / m g_F \mu_B B')^{1/3}$, with higher
angular momentum spin states residing outside those with lower
angular momentum.

Bose-Einstein condensates containing over $10^7$ $^{23}$Na atoms
were created in the $|1,-1\rangle$ state in a magnetic trap,
captured in the focus of an optical tweezers laser beam, and
transferred into an auxiliary ``science'' chamber as described in
Ref.~\cite{GCL02}.  In the science chamber, the condensate was
loaded into a microfabricated Ioffe-Pritchard magnetic trap formed
by a Z-shaped 50~$\mu$m~$\times$~10~$\mu$m electroplated copper
wire carrying current $I$ and an external magnetic bias field,
$B_\bot$, as described in Ref.~\cite{LCK02}. Typical wiretrap
parameters were $I = 720$~mA, $B_\bot = 5.3$~G, and $B_z = 1.3$~G,
resulting in a radial magnetic field gradient $B' = 180$~G/cm.
This produced axial and radial trap frequencies $\omega_z = 2 \pi
\times 4$~Hz and $\omega_\bot = 2 \pi \times 250$~Hz,
respectively.  Condensates in the wiretrap had $\geq 2 \times
10^6$ atoms, a Thomas-Fermi radius of $\approx 5\ \mu$m, and a
lifetime $\geq 25$~s~\cite{LSC02}.

\begin{figure*}
\includegraphics{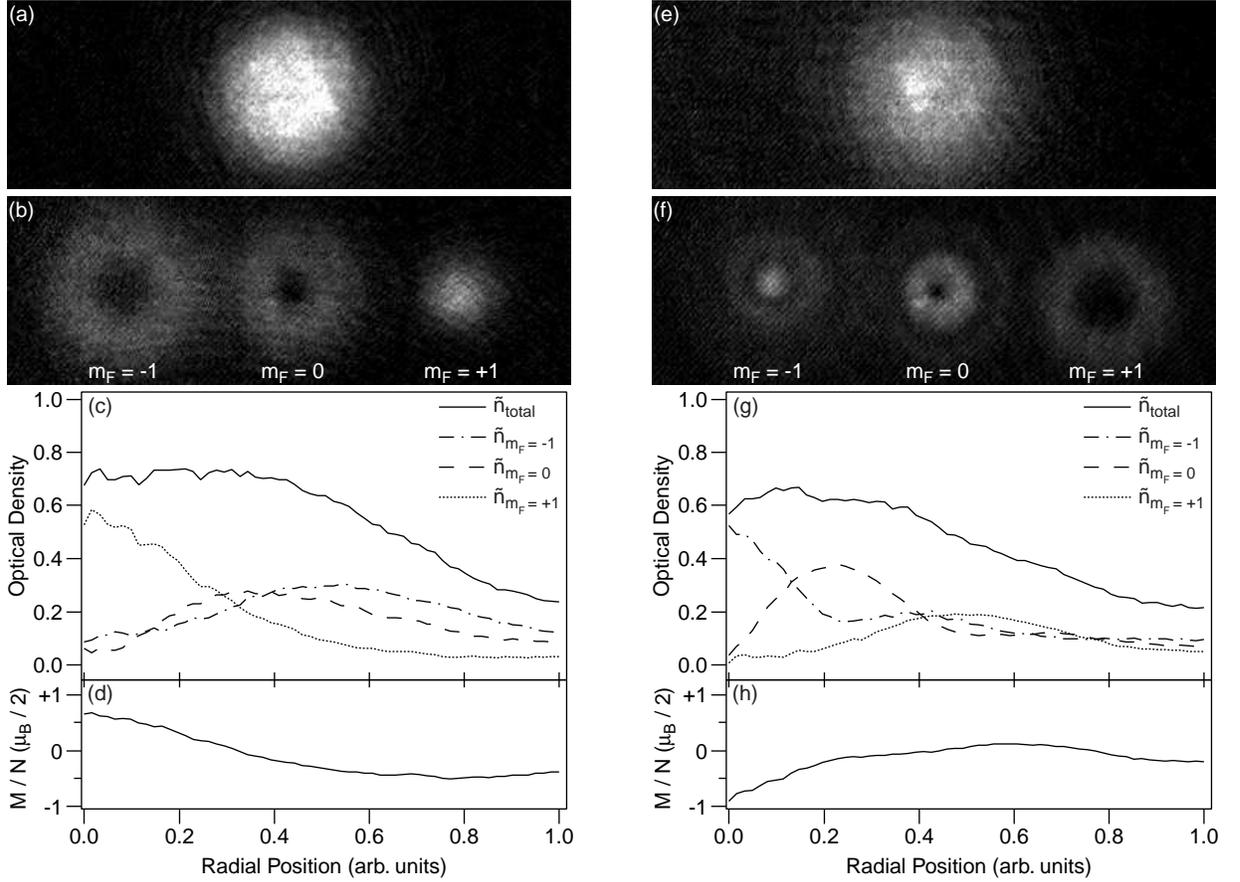}
\caption{Coreless vortex formation in a spinor Bose-Einstein
condensate. Coreless vortices were imprinted by ramping $B_z
\rightarrow 0$ and diagnosed by suddenly switching (a)-(d) $B_z
\ll 0$ and (e)-(h) $B_z \gg 0$. Axial absorption images display
the optical density of condensates after 20~ms of ballistic
expansion (a,e) without and (b,f) with a magnetic field gradient
applied to separate the different spin states. (c) and (g)
Azimuthally averaged optical density vs radial position for spin
components shown in (b) and (f), respectively. The radial
separation of the spin states resulted from their relative phase
windings and is a clear signature of the skyrmion/meron
wavefunction [Eq.~(\ref{e:texture})]. (d,h) Axial magnetization
per particle, $M/N = (\mu_B/2) \times (\tilde{n}_{m_F=+1} -
\tilde{n}_{m_F=-1})/\tilde{n}_{total}$, vs radial position. The
absorption imaging light was resonant with the $F=2 \rightarrow
F'=3$ transition. The atoms were optically pumped into the $F=2$
hyperfine level with a pulse resonant with the $F=1 \rightarrow
F'=2$ transition. This provided equal imaging sensitivity to each
spin state. The field of view in (a), (b), (e), and (f) is 1.0~mm
$\times$ 3.0~mm.\label{f:rings}}
\end{figure*}

Coreless vortices imprinted onto the condensate wavefunction by
adiabatically ramping $B_z \rightarrow 0$ are shown in
Fig.~\ref{f:rings}.  To observe the nature of the spin texture, an
axial bias field was switched on non-adiabatically along either
the negative [Fig.~\ref{f:rings}(a)-(d)] or positive
[Fig.~\ref{f:rings}(e)-(h)] z-axis.  Switching the axial bias
field on suddenly ``froze'' the atomic spins and effectively
``projected'' the condensate wavefunction onto a basis quantized
with respect to the local (axial) magnetic field.  This allowed
the spin states to be separated by a magnetic field gradient
applied during ballistic expansion. Switching the direction of the
axial projection field exchanged the roles of the $|1,-1\rangle$
and $|1,+1\rangle$ states. Figures~\ref{f:rings}(a) and
\ref{f:rings}(e) show the coreless nature of the vortices, while
Figs.~\ref{f:rings}(b) and \ref{f:rings}(f) show the concentric
cylinder structure resulting from the competition between the
atomic Zeeman energy and the kinetic energy of the rotating spin
states.  We assume that the two-dimensional ($\omega_z \ll
\omega_\bot$) ballistic expansion process simply magnifies the
condensate wavefunction, as it does in the expansion of a
single-component condensate with vortices~\cite{LPS98,DAM00}.

To imprint the coreless vortices, the axial bias field was ramped
linearly from $B_z = 1.3$~G to $B_z \approx 0$ in 10~ms. Along the
wiretrap axis, the magnetic field was
\begin{equation}
\label{e:curvature}
    \vec{B}(r=0,\phi,z) = (B_z + \frac{1}{2} B'' z^2)\ \hat{z},
\end{equation}
where quadratic terms neglected in Eq.~(\ref{e:ioffe}) are
included. The presence of axial magnetic field curvature, $B''
\approx 5$~G/cm$^2$, implies that the spin texture had a slight
axial dependence. Ramping $B_z \rightarrow 0$ quickly (10~ms)
compared to the axial trap period (250~ms) compressed the
condensate radially but left the axial extent of the condensate
unchanged.  Thus, the axial magnetic field variation, $\Delta
B_z$, was less than the radial magnetic field variation, $\Delta
B_r$, and the axial dependence of the spin texture was minimal.
With $B_z = 1.3$~G ($B_z \approx 0$~G), the condensate chemical
potential was $\mu \approx (\mu_B / 2) \times 3$~mG ($\mu \approx
(\mu_B/2) \times 27$~mG) yielding $\Delta B_z \approx 3$~mG
($\Delta B_r \approx 27$~mG). The images shown in
Fig.~\ref{f:rings} integrated the atomic number density along the
z-axis and therefore averaged over the minor axial variation to
the spin texture.

To project the condensate wavefunction onto a basis quantized with
respect to the local magnetic field, an axial bias field was
switched on at a rate of $\dot{B}_z = 2 \times 10^5$~G/s along
either the negative or positive z-axis.  For $0 \leq B_r \leq
\Delta B_r$, the Landau-Zener non-adiabatic transition
probability, $\exp (-\pi \mu_B B_r^2 / \hbar \dot{B}_z) \geq 0.9$,
was sufficiently close to unity that the atomic spins remained
``frozen'' during the sudden application of the axial bias field
and the spin texture could be accurately diagnosed. The magnetic
trap was switched off 100~$\mu$s after applying the axial bias
field allowing the atoms to expand ballistically. While the total
condensate density monotonically decreased as a function of radial
position [Figs.~\ref{f:rings}(a) and \ref{f:rings}(e)], the
density of each spin component peaked at a different radius
signifying a variation in the angular momentum per particle
between spin states [Figs.~\ref{f:rings}(b) and \ref{f:rings}(f)].

With the application of a projection field along the positive
z-axis, additional rings of atoms appeared in the $|1,-1\rangle$
and $|1,0\rangle$ states [Fig.~\ref{f:rings}(f)]. Possible
technical origins include the non-instantaneous turn on of the
projection field and the ballistic expansion process. Also,
ramping $B_z \rightarrow 0$ in 10~ms was non-adiabatic for atoms
near $r=0$ causing an atom loss of $\approx 50\%$ as observed in
previous work~\cite{LGC02}. If these atoms had not left the
condensate before the projection field was applied, they may have
contributed to the images displayed in Fig.~\ref{f:rings}(f). The
additional rings of atoms may also correspond to a low energy,
radial spin-wave excitation~\cite{HO98,OHM98}. However, we could
not identify any asymmetry between applying the projection field
along the positive versus negative z-axis that would account for
the presence of the extra rings in Fig.~\ref{f:rings}(f) but not
in Fig.~\ref{f:rings}(b).

Engineering topological states in a Bose-Einstein condensate has
received much theoretical
attention~\cite{INO00,WIH99,RUO00,MZS00,LEV00,RUA01}.  The
evolution of a condensate confined in a Ioffe-Pritchard magnetic
trap while ramping $B_z \rightarrow 0$ is described by a
position-dependent spin rotation about the $\hat{n}(\phi)$ axis
through an angle $\beta(r)$~\cite{LGC02}
\begin{equation}
\label{e:skyrmion}
    |\zeta(r,\phi,z)\rangle = e^{-i (\vec{\mathcal{F}}/\hbar) \cdot \hat{n}(\phi)
    \beta(r)} |\zeta_0\rangle,
\end{equation}
where $\vec{\mathcal{F}}$ is the spin operator and
$|\zeta_0\rangle = |F,m_z=m_F\rangle$ is a polarized spinor.  For
$|\zeta_0\rangle = |1,-1\rangle$, Eq.~(\ref{e:skyrmion}) gives the
condensate spinor in the laboratory frame as
\begin{equation}
\label{e:texture}
    |\zeta(r,\phi,z)\rangle =   \begin{array}{l}
                                \cos^2 (\beta(r)/2)\ |1,-1\rangle\ + \\
                                \frac{-1}{\sqrt{2}} \sin (\beta(r))\ e^{i \phi}\ |1,0\rangle\ + \\
                                \sin^2 (\beta(r)/2)\ e^{i 2 \phi}\ |1,+1\rangle .
                                \end{array}
\end{equation}
$\beta(0) = 0$ and skyrmions (merons) are described by the
boundary condition $\beta(\infty) = \pi$ ($\beta(\infty) =
\pi/2$).

The radial dependence of $\beta(r)$ is determined by requiring the
Gross-Pitaevskii energy functional,
\begin{equation}
\label{e:energy}
    E = \int d^3\vec{r}\ n \left( \frac{\hbar^2}{2m}\langle\nabla\zeta|\nabla\zeta\rangle + V + \left(\frac{c_0}{2}+\frac{c_2}{2} \frac{|\vec{F}|^2}{\hbar^2} \right)n \right),
\end{equation}
be stationary with respect to variations in $\beta$.  In writing
Eq.~(\ref{e:energy}) we have made the Thomas-Fermi approximation,
$\langle\nabla\zeta|\nabla\zeta\rangle = \sum_{m_z=-F}^F
\nabla\zeta_{m_z}^\dagger \cdot \nabla\zeta_{m_z}$, $V = -g_F
\mu_B \vec{F} \cdot \vec{B} / \hbar$, $\vec{F} = \langle \zeta |
\vec{\mathcal{F}} | \zeta \rangle$, $c_0 = 4\pi\hbar^2\bar{a}/m$,
and $c_2 = 4\pi\hbar^2\Delta a / m$.  Here $\bar{a} =
(2a_0+a_2)/3$ and $\Delta a = (a_2-a_0)/3$ characterize two-body
interactions, where $a_0$ and $a_2$ are scattering lengths for
collisions with total angular momentum $F=0$ and $F=2$,
respectively~\cite{HO98}. For $a_2 > 0$ the atomic interactions
are anti-ferromagnetic (polar), as in $^{23}$Na~\cite{SIS98},
while for $a_2 < 0$ they are ferromagnetic, as in
$^{87}$Rb~\cite{KBG01}.

Using Eq.~(\ref{e:texture}) as a trial spinor, we find that
$\beta(r)$ satisfies the meron boundary conditions and varies from
$\beta(0) = 0$, due to the kinetic energy of the spin texture, to
$\beta(\infty) = \pi/2$, due to the atomic Zeeman energy, over a
characteristic length scale given by the larger of $(\hbar^2/m g_F
\mu_B B')^{1/3}$ and $|B_z|/B'$, in reasonable agreement with the
results presented in Fig.~\ref{f:rings}.  We observed that the
boundary condition $\beta(0) = 0$ was maintained regardless of the
sign of $B_z$, i.e.\ atomic spins along the trap axis always
remained in the initial state. This emphasizes the non-trivial
topology of the spin texture.  If the atomic spins had simply
followed the local magnetic field, $\beta(r)$ would satisfy $\tan
\beta(r) = B'r/B_z$.  Thus, scanning $B_z$ from slightly positive
to slightly negative would instantaneous change the boundary
condition at the origin from $\beta(0) = 0$ to $\beta(0) = \pi$
and flip the atomic spins along the trap axis.

The trial spinor in Eq.~(\ref{e:texture}) does not have the most
general form since it was derived by rotating a polarized spinor
[Eq.~(\ref{e:skyrmion})] and is inherently ferromagnetic,
$|\vec{F}| = \hbar$.  Accordingly, the spin-dependent interaction
term in Eq.~(\ref{e:energy}) does not vary with $\beta$ and
therefore does not contribute to the determination of $\beta(r)$.
This restriction is not severe since the Zeeman energy dominates
the spin-dependent interaction energy in our experiment.

It should be possible to overlap the condensate with an optical
dipole trap so that, after ramping $B_z \rightarrow 0$, relaxing
the radial magnetic field gradient would allow for the
spin-dependent interactions to determine the evolution of the
condensate. It is predicted that skyrmions/merons in condensates
with anti-ferromagnetic (ferromagnetic) interactions are unstable
(stable)~\cite{KHS01nature,KHS01pra,BCS02,MMK02prl,ISM02,ZLL02,MCS02pra,MMK02pra}.
At present, excitations created during the field ramping process
have prevented us from characterizing the stability of coreless
vortices in an anti-ferromagnetic $^{23}$Na condensate. However,
we were able to imprint the spin texture in the presence of the
optical dipole trap, as well as reproduce the results of
Ref.~\cite{LGC02} by fully inverting the axial bias field to
produce vortices with a $4\pi$ phase winding in an optical dipole
trap. At zero magnetic field, it may then be possible to observe
multiply charged vortices in a spinor condensate ``unwind''
themselves as predicted in Ref.~\cite{HO98}. The unwinding process
is precisely the position-dependent spin rotation demonstrated in
previous work~\cite{LGC02}, with the spin texture investigated
here as an intermediate state~\cite{MCS02pra}.

In conclusion, we have demonstrated a robust technique for
creating coreless vortices in a Bose-Einstein condensate. Our
technique can be extended to generate spin textures with arbitrary
winding number and variable angular momentum per particle by using
higher-order, axisymmetric multipole magnetic fields and
condensates with different spin. This work opens up the
opportunity to study the stability of topological defects in
spinor Bose-Einstein condensates.

We are grateful to K.~Machida for bringing the spin texture
studied here to our attention.  We thank T.~Pasquini for
experimental assistance and J.R.~Anglin and M.~Crescimanno for
valuable discussions. This work was funded by ONR, NSF, ARO, NASA,
and the David and Lucile Packard Foundation.

\end{document}